\documentclass[preprint,12pt,authoryear,5p]{elsarticle}

\usepackage{amssymb}
\usepackage{amsmath}
\usepackage{comment}
\usepackage{booktabs}
\usepackage{mathtools}
\usepackage{hyperref}
\usepackage{tabularx}
\usepackage{multirow} 
\usepackage{adjustbox}
\usepackage{blindtext}
\usepackage{float}
\usepackage{graphicx}
\usepackage{subfig}
\usepackage{phfcc}
\usepackage{bm}
\usepackage[
    starfontserif 
    ]{starfont}
\usepackage{natbib}

\DeclareSymbolFont{starfontsym}{OT1}{sts}{m}{n}

\DeclareMathSymbol{\mathVenus}{\mathord}{starfontsym}{103}
\DeclareMathSymbol{\mathEarth}{\mathord}{starfontsym}{76}
\DeclarePairedDelimiter\abs{\lvert}{\rvert}%

\UseRawInputEncoding

\journal{Acta Astronautica}

\begin{document}

\begin{frontmatter}

\title{Italian Spring Accelerometer measurements of unexpected Non Gravitational Perturbation during BepiColombo second Venus swing-by }

\author[inst1]{Carmelo Magnafico}

\affiliation[inst1]{organisation={IAPS - Institute of Astrophysics and Space Planetology / INAF National Institute for Astrophysics},
            addressline={Via del Fosso del Cavaliere, 100}, 
            city={Rome},
            postcode={00133}, 
            state={Lazio},
            country={Italy}}

\author[inst1]{Umberto De Filippis}
\author[inst1]{Francesco Santoli}
\author[inst1]{Carlo Lefevre}
\author[inst1]{Marco Lucente}
\author[inst1]{David Lucchesi}
\author[inst1]{Emiliano Fiorenza}
\author[inst1]{Roberto Peron}
\author[inst1]{Valerio Iafolla}

\begin{abstract}

The Italian Spring Accelerometer (ISA) is a three axis mass-spring accelerometer, one of the  payloads of the BepiColombo joint space mission between the European Space Agency (ESA) and the Japan Aerospace Exploration Agency (JAXA).
At launch in October 2018, BepiColombo started its seven-year cruise as a stack of three different modules, overall named Mercury Composite Spacecraft (MCS). The spacecraft will provide BepiColombo the necessary Delta V to reach Mercury with its electric thrusters and along with one, two and six gravity assists, respectively with Earth, Venus and Mercury. 
The accelerometer is accommodated on the Mercury Planetary Orbiter (MPO) module and, jointly with the Ka-band Transponder (KaT) tracking data, will primarily serve the BepiColombo Radio Science Experiment (BC-RSE). 
During the second Venus swing-by, strong tidal effect and external perturbations was expected to act on the spacecraft and to become detectable by ISA. The swing-by had a closest approach of about 550 km and the gravity gradient expected on the IDA sensing elements was perfectly measured. Hence, in this paper, the first direct Gravity Gradient effect detection generated by an extraterrestrial body is shown. Nevertheless, around the closest approach, the measurements evidenced a spurious acceleration event lasting for several minutes. This work, exploiting  information on the Attitude and Orbit Control System (AOCS) commanded torques, focuses and analyses this ISA acceleration signal, ascribing it to a net force really acting on the MCS spacecraft. Furthermore, using an estimation method, the application point of the force is confined to an area close to the MPO radiator.

\end{abstract}




\end{frontmatter}

\emergencystretch 3em

\section{The BepiColombo Mission to Mercury}
\label{Intro}

\subsection{The BepiColombo Mission}
\label{BC_mission_intro}
Planet Mercury, the closest planet to the Sun, (0.30 AU - 0.46 AU), orbits in an environment exposed to intense particle's fluxes and to a solar constant ranging from approximately 6,000 to 14,000 $W/m^2$, which is up to ten times that at Earth. Therefore, reaching Mercury with a spacecraft needs a complex design, accurate operation planning and years of navigation to win the gravitational potential gap. Is not the case that to date, the number of spacecraft that could reach Mercury were only two, both by NASA: Mariner 10 in 1974-75 made three swing-bys, while MESSENGER (MErcury Surface, Space ENvironment, GEochemistry and Ranging) in 2011 \cite{Messenger_2006AdSpR..38..564M}, was the unique to enter in a stable hermean orbit. As a comparison, nowadays, the overall Mars exploration missions are 52.\\
Nevertheless, Mercury is still a key scientific element for the understanding of Solar System evolution. The knowledge of its interior, the presence of a significant magnetosphere, the nature of some surface features (such as the hollows and the clusters of rimless depressions) are still open points for the Mercury science. To this aim, already in 2000, and then in 2008 with a complete review of size, targets and budget, the BepiColombo mission was approved as joint mission between the European Space Agency (ESA) and the Japan Aerospace Exploration Agency (JAXA) \cite{Benkhoff2010BepiColombo-ComprehensiveGoals}. 

The mission is dedicated to Giuseppe "Bepi" Colombo (1920-1984), a Paduan mathematician and engineer, who specialised in celestial mechanics. He discovered the Mercury's spin-orbit 3:2 resonance and contributed to the calculation of the NASA Mariner 10 probe's orbit, by suggesting to exploit this peculiarity to carry out three fly-bys at Mercury instead of one alone.\\
The BepiColombo mission hosts 16 payloads and aims at hundreds of scientific goals for a complete characterisation of the planet's interior, its surface, the exosphere, and the magnetosphere and to test Einstein's theory of General Relativity. To fulfil those ambitious scientific goals, the overall assembly is designed in a modular configuration of four parts: the Mercury Transfer Module (MTM), the Mercury Planetary Orbiter (MPO),  the Mercury Magnetosphere Orbiter (MMO or Mio) and the Magnetospheric Orbiter Sunshield and Interface Structure (MOSIF). 
\begin{figure}[!ht]\centering
    \subfloat[]{
        \label{fig:MCS_configuration_a}
        \includegraphics[width=1\linewidth]{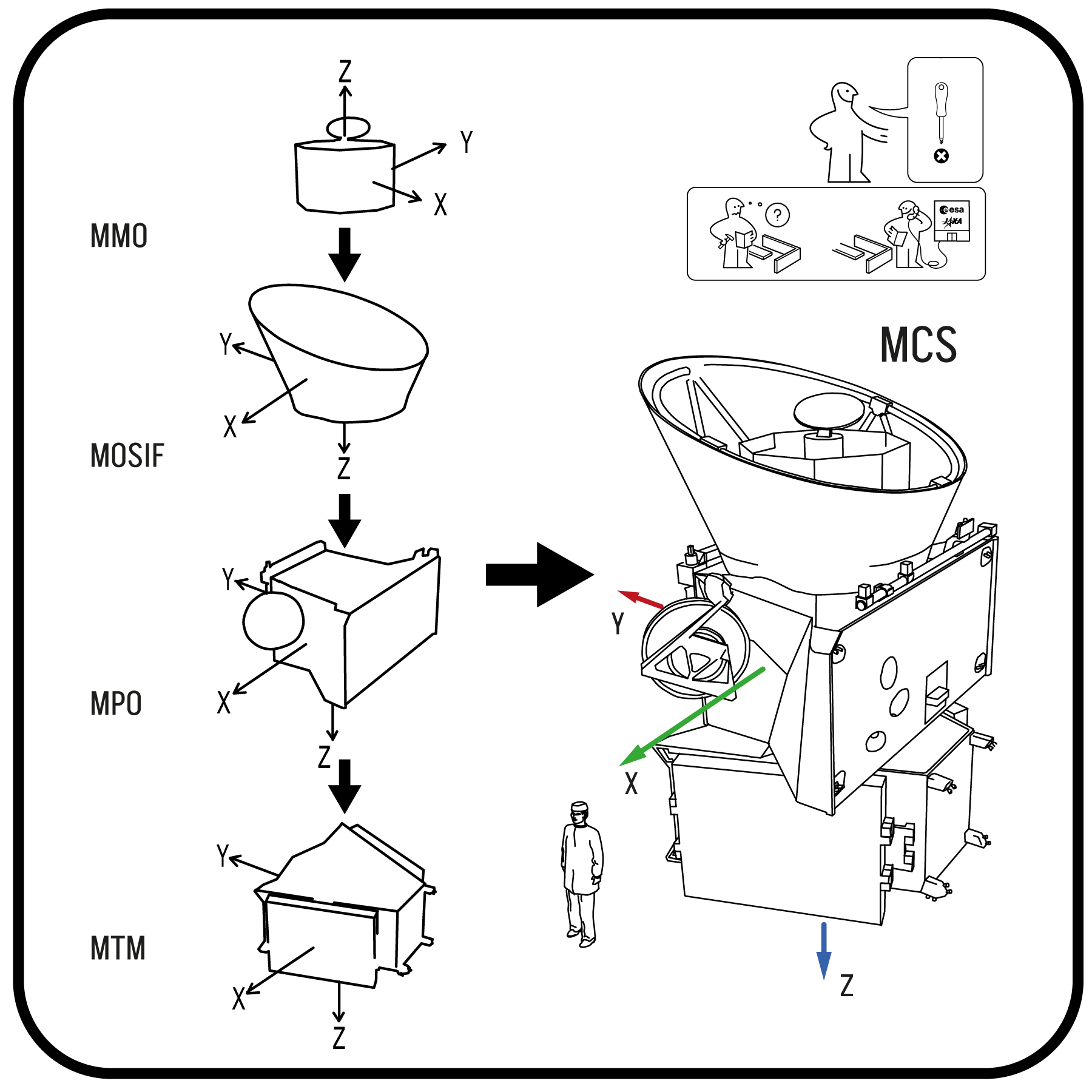}
    }\par
    \subfloat[]{
        \label{fig:MCS_configuration_b}
        \includegraphics[width=.45\linewidth]{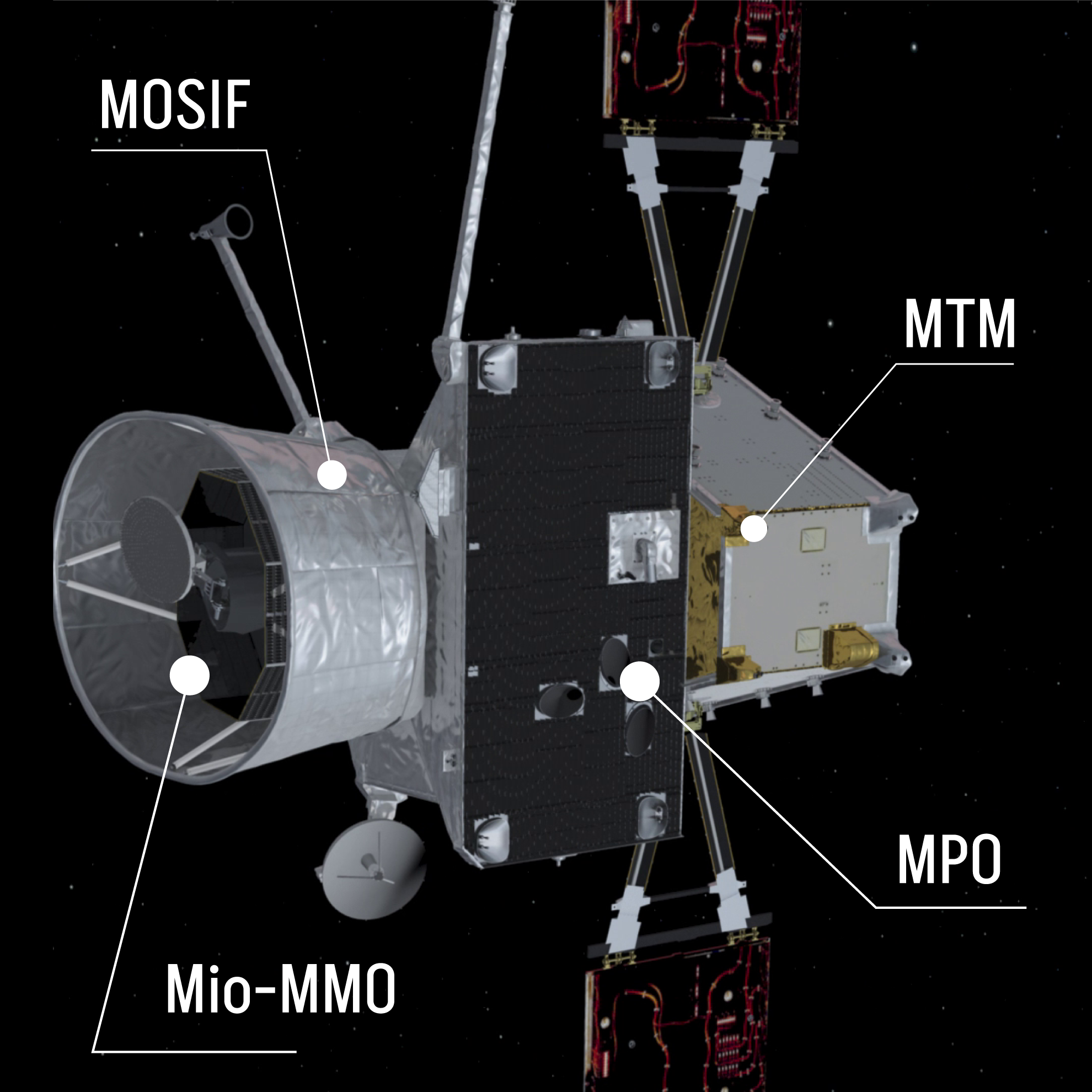}
    }\hfill
    \subfloat[]{
        \label{fig:MCS_configuration_c}
        \includegraphics[width=.45\linewidth]{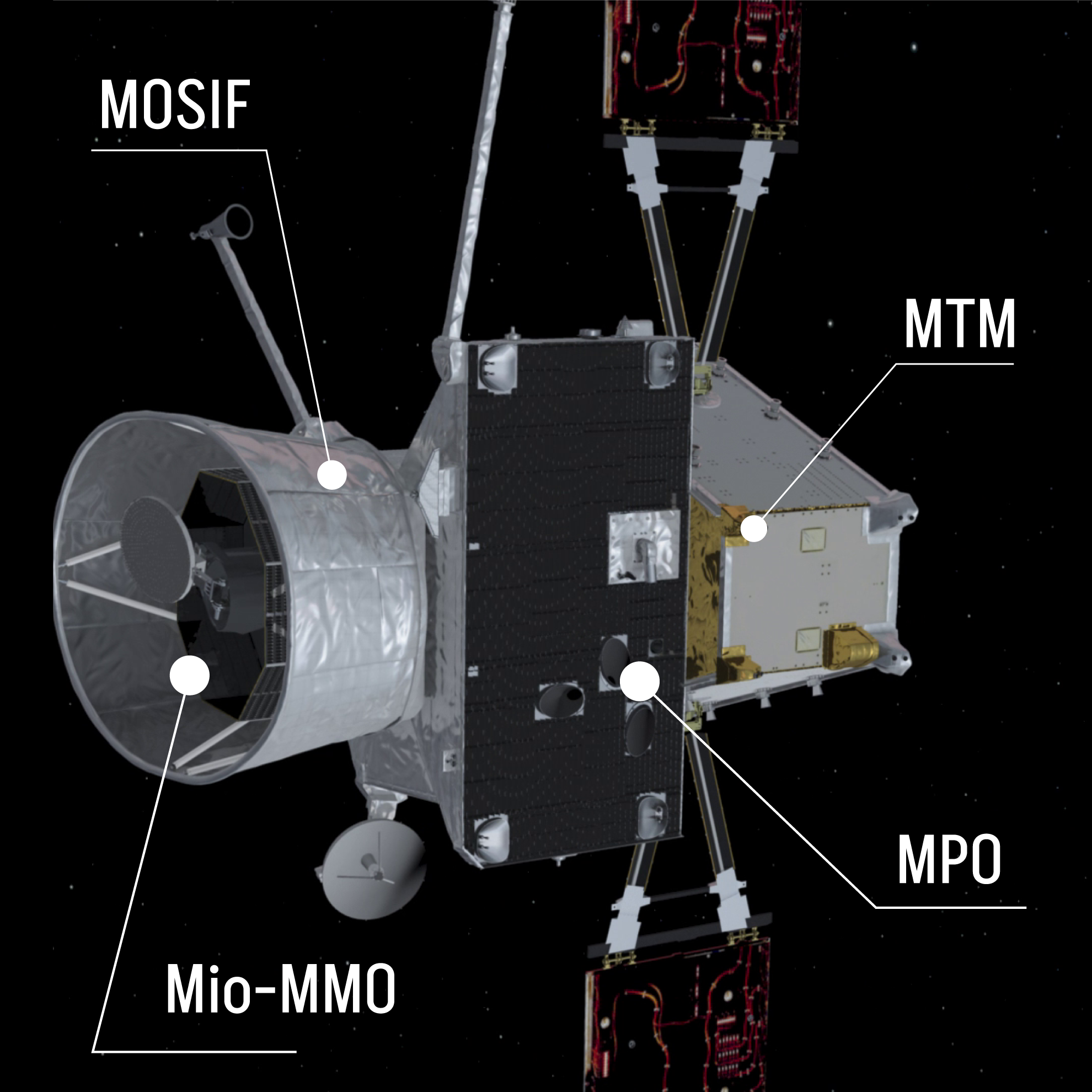}
    }
    \caption{BepiColombo Mercury Composite Spacecraft (MCS). a) A diagram of BepiColombo at launch b) MCS in cruise configuration, backward view. c) MCS in cruise configuration, upward view. }
    \label{fig:MCS_configuration}
\end{figure}

\begin{table*}[t]
    \centering
    \caption{MPO and Mio Payloads list}
    \label{Tab:BepiColombo_Payloads_list}
    \begin{adjustbox}{width=\textwidth}
        \begin{tabular}{l l l}
        \toprule
        \multicolumn{3}{c}{BepiColombo Payloads list} \\
        \toprule \hline
        $\textbf{Name}$ & $\textbf{Type}$ & $\textbf{position}$\\
        \midrule
        BELA&Laser Altimeter&MPO\\
        ISA&Three Axis Accelerometer&MPO\\
        MERMAG&Magnetometer&MPO\\
        MERTIS-TIS&IR spectrometer&MPO\\
        MGNS&Gamma ray and neutron spectrometer&MPO\\
        MIXS&X-ray spectrometer&MPO\\
        Ka-T&Radio science Ka-band transponder&MPO\\
        PHEBUS&UV spectrometer&MPO\\
        SERENA&Neutral and ionised particle analyser&MPO\\
        SIMBIO-SYS&High resolution + stereo camerasVisual and NIR spectrometer&MPO\\
        SIXS&Solar monitor&MPO\\
         \midrule
        MDM&Dust Monitor&Mio\\
        MGF&Magnetometer&Mio\\
        MPPE&Plasma, high-energy particle and energetic neutral atom detector&Mio\\
        MSASI&Mercury Sodium Atmospheric Spectral Imager&Mio\\
        PWI&Sets of Electric Field sensors&Mio\\
        \bottomrule \hline
        \end{tabular}
    \end{adjustbox}
\end{table*}

Launched on the 20th of October 2018 from French Guiana spaceport on board the VA-245 Ariane 5 ECA 5105, the spacecraft is currently facing the cruise phase in the Mercury Composite Spacecraft (MCS) configuration, as shown in Figure (\ref{fig:MCS_configuration}). The whole MCS is propelled by four, max thrust 145 mN each, solar electric thrusters mounted on board the MTM module, also housing 40 $m^2$ solar panels, 10 meters tip-to-tip long, able to generate up to 13,000 $W$. The arrival at Mercury is foreseen in December 2025, after about seven years in the interplanetary space on a trajectory designed to exploit $ \Delta V $ from one flyby with the Earth, two with Venus and six with Mercury. At the end of the last electric thrust arc, the MCS will detach the MTM module and the MPO chemical propellers will start the Mercury Orbit Injection (MOI). MPO and Mio will move to high eccentricity $ 11500\ km$ apoherm, $ 400\ km$ periherm, polar orbit, that will be the Mio final orbit. Finally, after the Mio/MPO separation, MPO will reduce its orbit eccentricity reaching $1500\ km$ apoherm - $ 400\ km$ periherm orbit. 
The Mio spacecraft with its five payloads (Table \ref{Tab:BepiColombo_Payloads_list}) is dedicated to the measurement of the exosphere and magnetosphere of the Mercury environment. At the contrary, the MPO will focus on the planetary complete characterisation, using payloads (Table \ref{Tab:BepiColombo_Payloads_list}) dedicated to imaging, radio tracking, spectrography, laser-altimetry and magnetospheric measurements.\\ 
In particular, the Mercury Orbiter Radioscience Experiment (MORE) will have a key role in gravity field estimation, in characterisation of the planet's interior status and the estimation of post-Newtonian parameters. MORE will use the radio tracking Ka-T three links tracking data and the Italian Spring Accelerometer (ISA) measurements.

\subsection{The ISA payload}
\label{ISA}
The Italian Spring Accelerometer is a three-axis high-sensitivity accelerometer (one-dimensional sensing elements per acceleration axis forming a quasi-independent orthogonal triplet). The ISA's final scope is to measure the three components of the non-gravitational acceleration vector acting on the spacecraft body, \cite{Santoli2020ISASpace}. 
ISA measurement band extends from $3\times10^{-5}$ Hz to $10^{-1}$ Hz. Hence, its measurements shall be considered as relative and not absolute accelerations.\par
The main expected accelerations measured by ISA can be divided into non-conservative and conservative accelerations. So, respectively, accelerations that can cause a net move of S/C CoM and accelerations causing only oscillation of CoM from its nominal trajectory. 
The non-conservative accelerations are as follows:
\begin{itemize}
    \item The direct solar radiation pressure (SRP)
    \item The visible albedo radiation pressure
    \item The planetary thermal emission radiation pressure
    \item The S/C anisotropic thermal re-emission
    \item Outgassing 
    \item Thrusters or tanks gas leaks
\end{itemize}
On the other hand, the conservative accelerations are: 
\begin{itemize}
    \item Inertial rotations accelerations 
    \item Gravity Gradient effects
    \item Appendices movements
    \item S/C Internal parts low band vibrations.
\end{itemize}

Since ISA is sensitive to Non-Gravitational Perturbations (NGPs), a relevant part of ISA science data are foreseen to be used in the context of the BepiColombo Radio Science Experiment (BC-RSE) to perform at best the Precise Orbit Determination (POD). The aim of the instrument is to continuously measure the accelerations acting on-board both in visibility of the ground-station and in out-of-pass conditions \cite{U_DeFilippis_2024JGCD...47..685D}.\par
\begin{figure}[hbt!]\centering
    \subfloat[]{
        \label{fig:ISA_sensing_mass_a}
        \includegraphics[width=.75\linewidth]{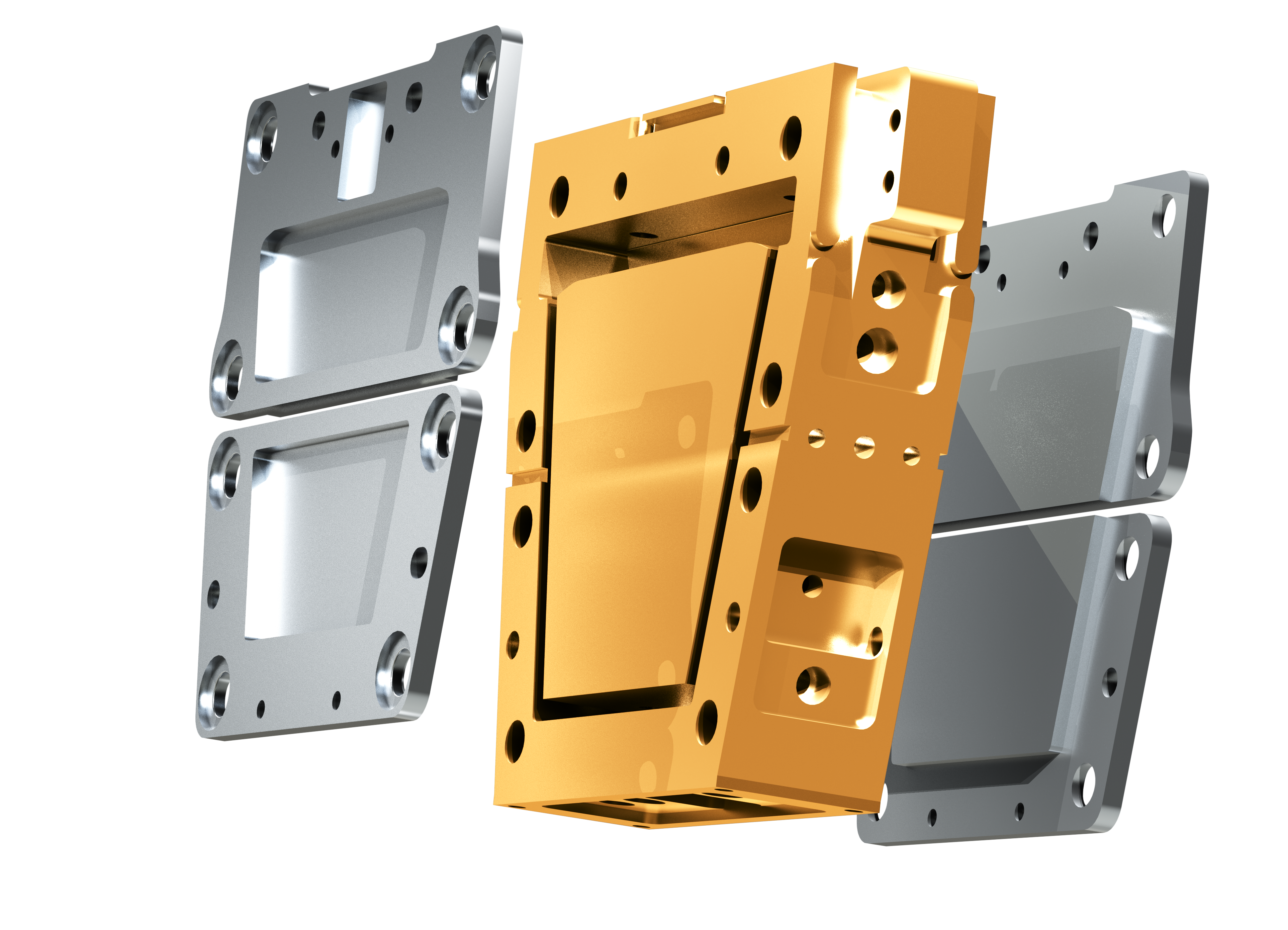}
    }\par
    \subfloat[]{
        \label{fig:ISA_sensing_mass_b}
        \includegraphics[width=.75\linewidth]{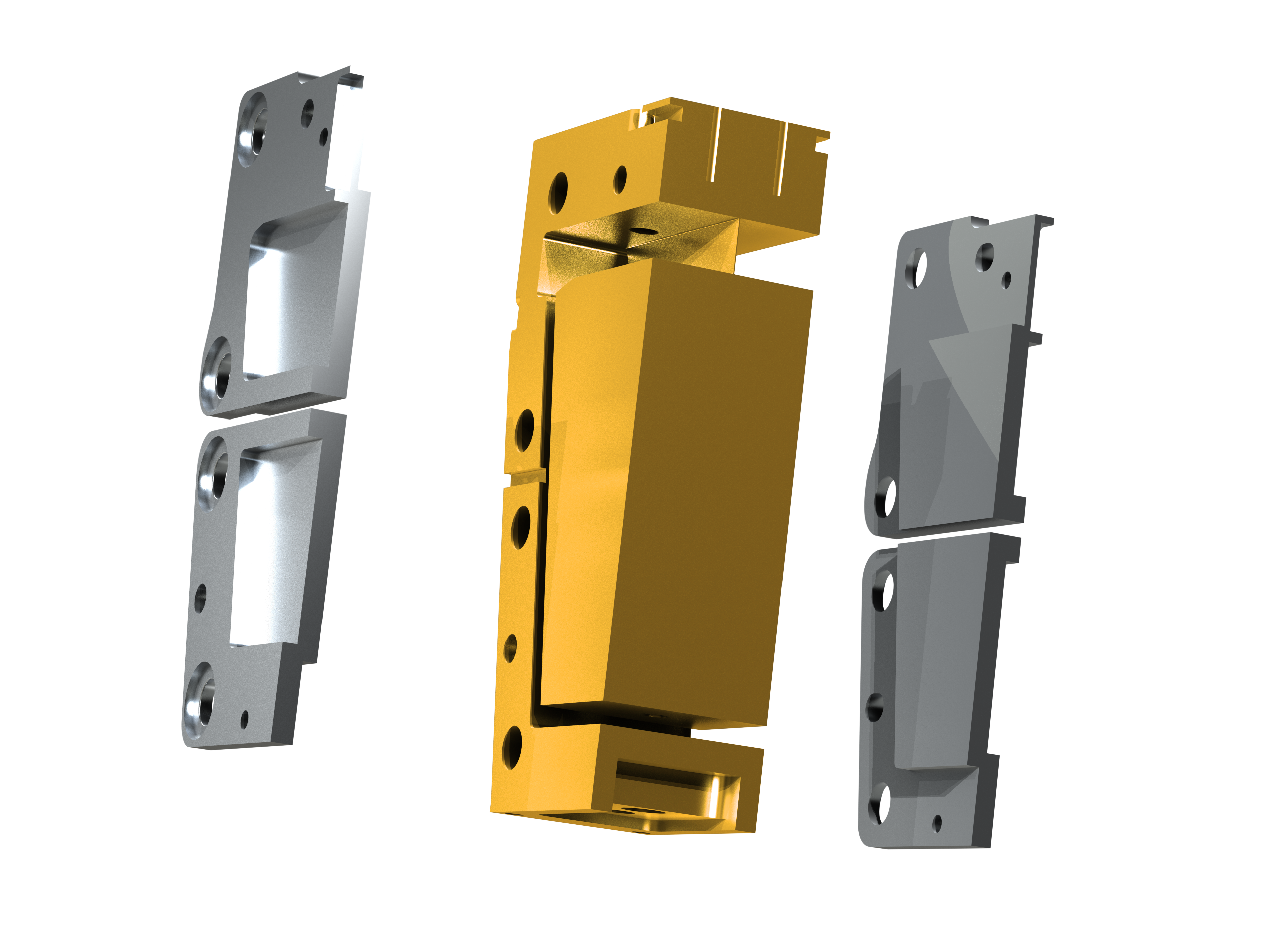}
    }
    \caption{Exploded view of one of the three ISA Sensing Element. a) The trapezoidal sensing moving mass b) Section of mass evidencing the blade shape spring. }
    \label{fig:ISA_sensing_mass}
\end{figure}

ISA is made of three sensing elements, designed as mechanical harmonic oscillators. Each one consists of a proof mass hanged from an external rigid frame through a flexural element (spring) with low elastic constant. The sensing elements are equipped with two pairs of pick-up and actuation plates. The pick-up plates are used to read the acceleration signal by capacitive sensing of the movement of the central mass. On the other hand, the actuation plates are used to generate electrostatic forces to centre the sensing mass with respect to the electrical zero of pick-up plates and to internally calibrate each element by acting on the mass with a known on-ground calibrated signal. (figure (\ref{fig:ISA_sensing_mass})).

\begin{figure}[!ht]\centering
    \subfloat[]{
        \label{fig:IDAassembly_a}
        \includegraphics[width=.9\linewidth]{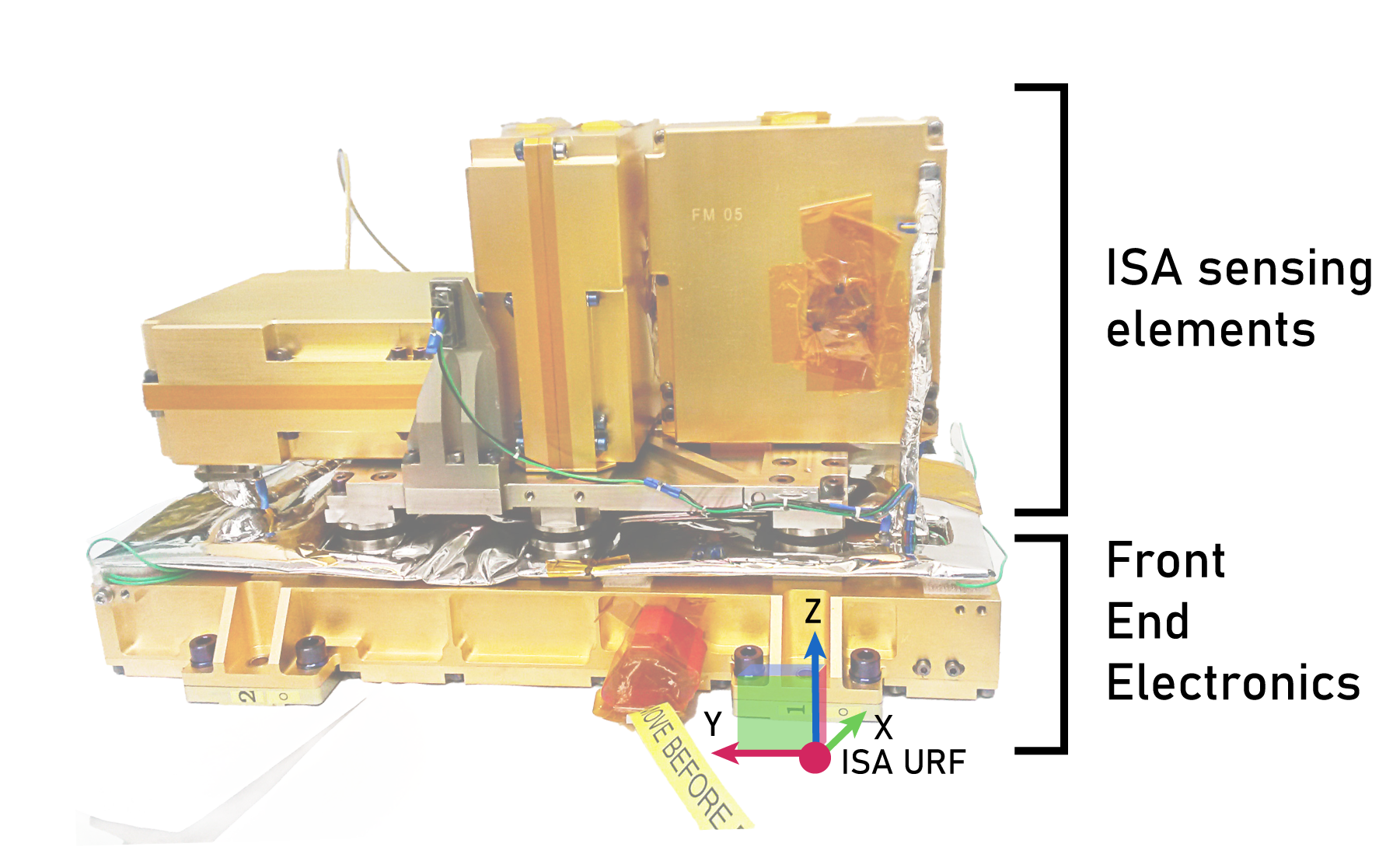}
    }\par
    \subfloat[]{
        \label{fig:IDAassembly_b}
        \includegraphics[width=.9\linewidth]{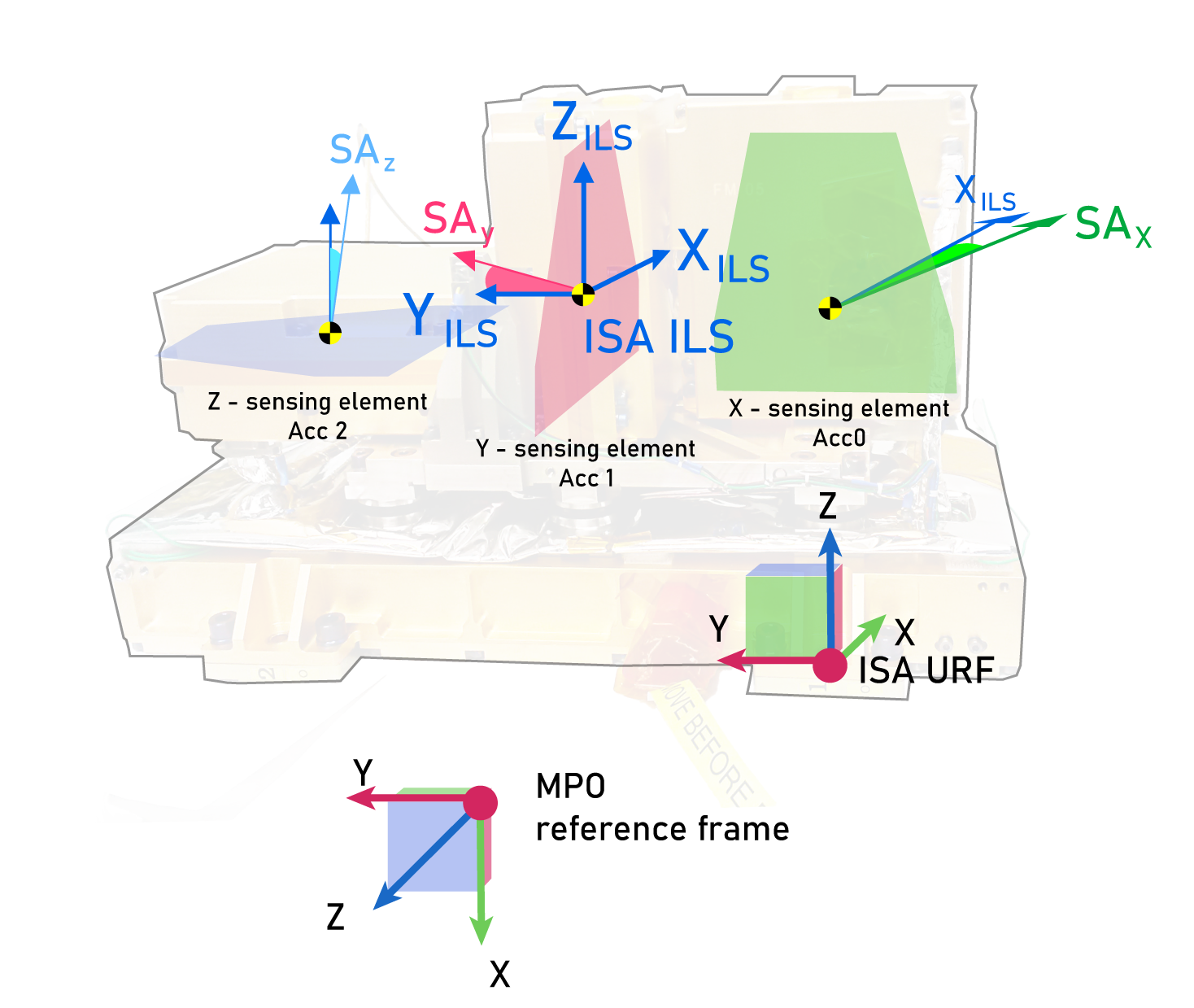}
    }
    \caption{ ISA - IDA assembly. Figure a) IDA Flight Model assembly. Figure b) ISA ILS internal reference frames as placed on the assembly.}
    \label{fig:IDAassembly}
\end{figure}

In order to better take under control electrical and thermal disturbances, it has been decided to decouple as much as possible the digital electronics and the power supply section from the detectors. This results in a configuration that foresees two independent boxes. Therefore, the ISA instrument configuration is based on two units: the ISA Detector Assembly (IDA) and the ISA Control Electronics (ICE) that interfaces the MPO for power supply and communication. 
IDA is made up by a Front End Electronics (FEE), placed at the bottom of the box, and by the three sensing elements assemblies, each enclosed in dedicated shields, in order to decouple as much as possible the sensing elements from the temperature environmental disturbances. At the same time, shields reduce the power consumption of the active thermal control system. The configuration is described in figure (\ref{fig:IDAassembly_a}).

As depicted in figure (\ref{fig:IDAassembly_b}) accelerometers Center of Mass (CoM) are not located in same point and each sensing element has a Sensing Axis (SA), identified during the on-ground calibration as the direction in which a known acceleration external to the assembly gives the ma\-ximum signal. Hence, the three sensing elements SA, despite designed to point in the three space directions, does not represents an ideal orthogonal tern. To fulfill the need to address the measuring acceleration to a unique, fixed point on the assembly, reference frame, the ISA Instrument Line of Sight (ISA\_ILS) is introduced. ISA\_ILS is defined as the closer orthogonal tern to the non-orthogonal reference formed by the three sensing axes, and it is placed in the Y sensing element CoM. ISA\_ILS is also called ISA vertex, as it represents the MPO on-board vertex with respect to what the MORE experiment will perform the Precise Orbit Determination.   
The operation of converting an acceleration measured by ISA sensing elements in a ISA ILS acceleration is not a simple base rotation. Indeed, positional accelerations differences as Gravity Gradient effect and Inertial rotations accelerations between the sensing elements shall be kept into account. The calculation is done on-ground by ISA pipelines and is called "vertex reduction".\\
The attitude of ISA with respect to the spacecraft is depicted in figure(\ref{fig:IDAassembly_b}). The figure summarise how MPO reference frame (MPO\_RF) is oriented with respect to the ISA\_URF and ISA\_ILS. The relation between  ISA\_URF and MPO\_RF is then a very simple rotation, since, by design, the two reference frames are 90° anticlockwise Y-rotated, so the $X_{ILS} \cong -Z_{MPO}$, $Y_{ILS} \cong Y_{MPO}$ and $Z_{ILS} \cong X_{MPO}$.

\subsection{The Cruise Phase}
\begin{figure}[!t]
\centering
    \includegraphics[width=0.5\textwidth]{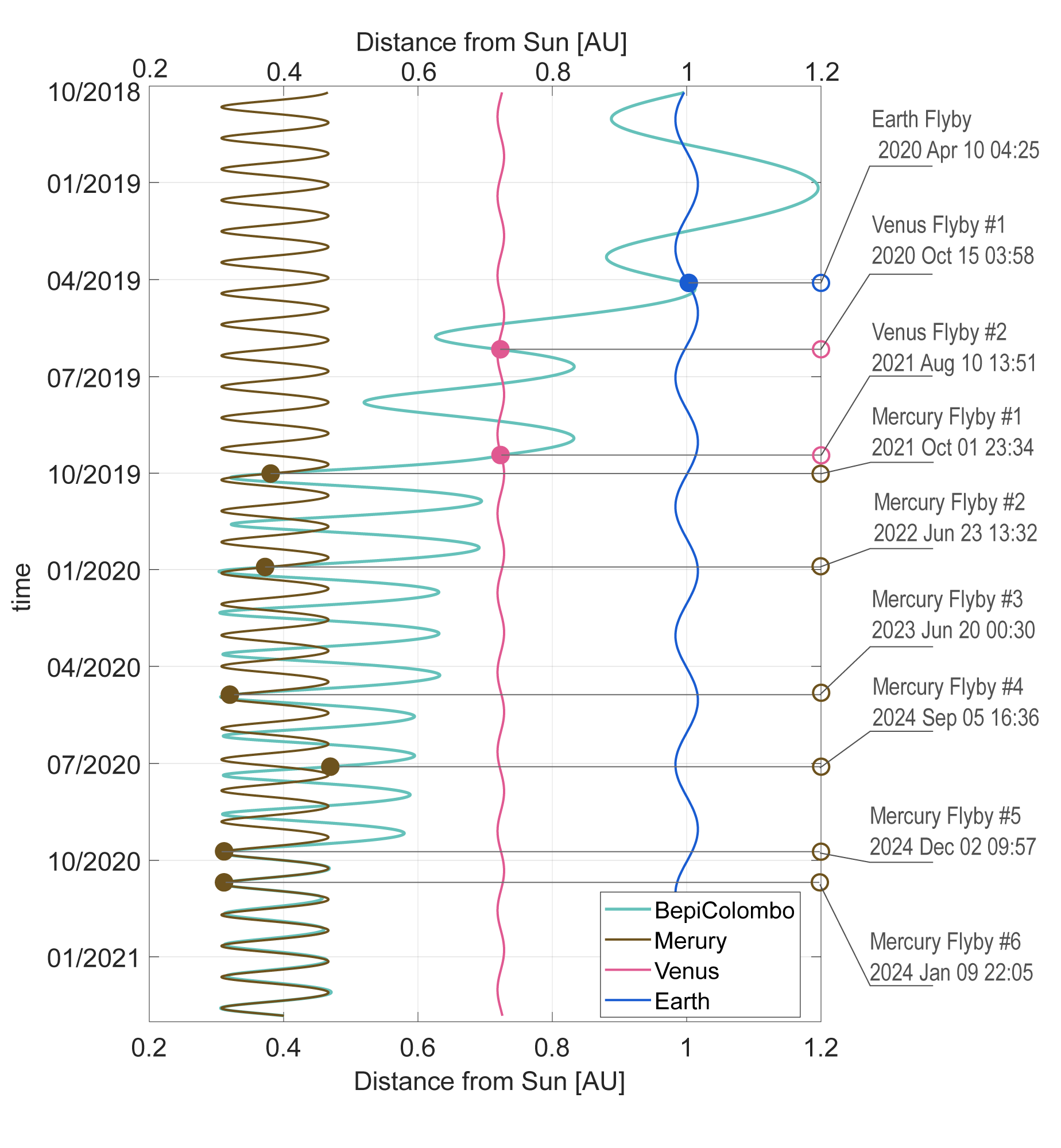}
    \caption{BepiColombo flybys history in terms of event times and distance from Sun.}
    \label{fig:BepiFlyby}
\end{figure}

The Mercury position in the Solar System is so close to the Sun that $\Delta V$ needed for a simple Hohmann transfer from the Earth is comparable to one a probe would use to reach Jupiter.  
For that reason the BepiColombo cruise trajectory towards Mercury implies a complex set of operations to minimise the $\Delta V$ and so the propellant stored on board. During the cruise, pe\-riods of thrusted arcs using the MTM SEPS (Solar Electric Propulsion System) alternate with periods of coasting arcs, in which the spacecraft is in pure free fall around the Sun. Moreover, the cruise foresees several planetary flybys, summarised in Figure \ref{fig:BepiFlyby}.\par
The BepiColombo cruise evolution was considered so crucial for the entire mission success that, at launch, the only scientific activity foreseen during the cruise was the MORE Superior Solar Conjunction Experiment (SCE) with the ISA support. Therefore, when approaching the Earth Flyby in 2020, the Mission Manager and the System Operation Manager relaxed those specifications considering the flybys as an important opportunity for payloads able to be switched-on despite the MCS configuration, to collect data for characterise other planet environments or calibrate instruments for the nominal Hermean mission.   \cite{Mangano2021BepiColomboMercury}

\subsection{The Venus swing-by \#2 (VSB2)}
\label{BC_VSB_2}
\begin{figure}[!ht]
\centering
    \includegraphics[width=0.5\textwidth]{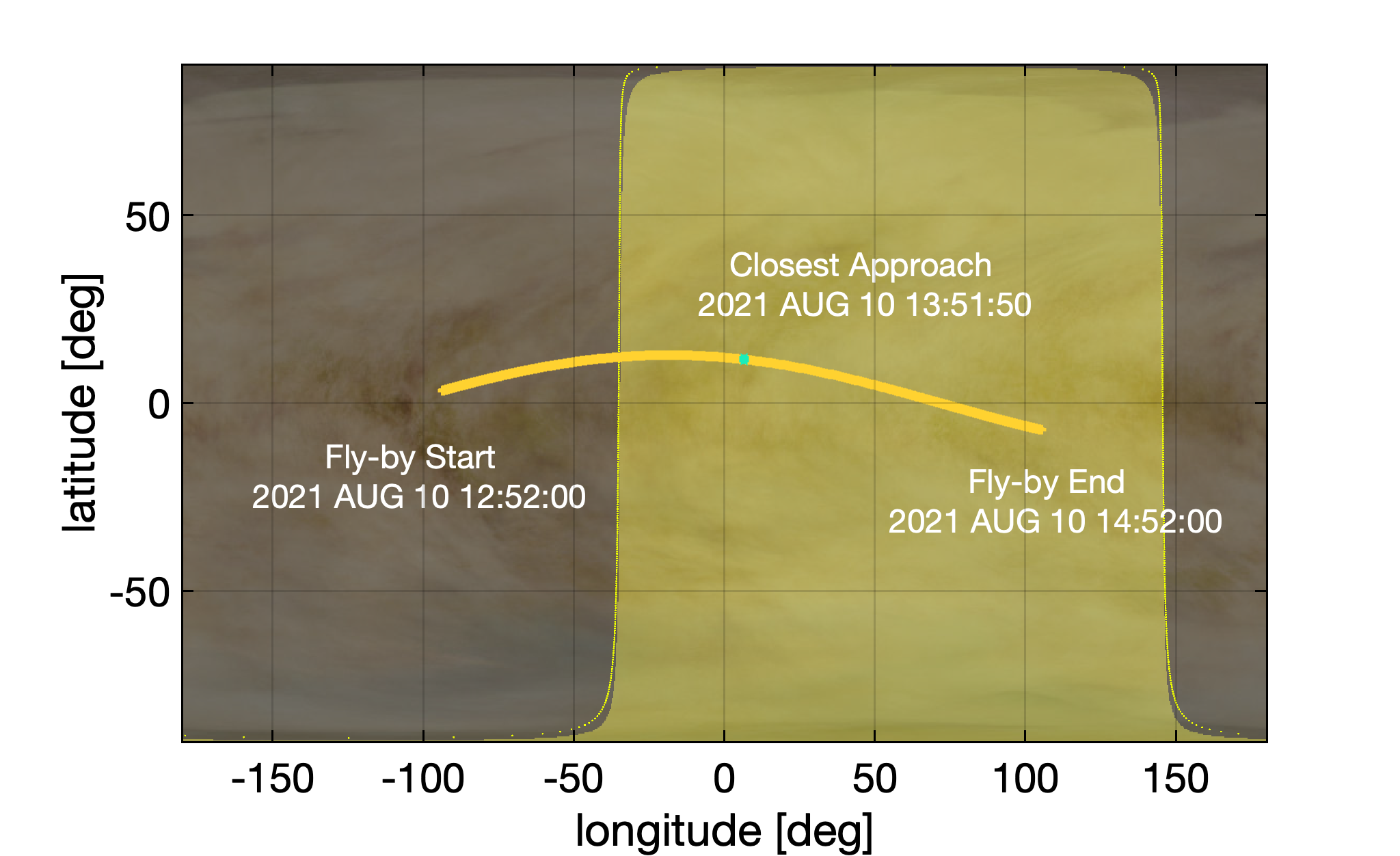}
    \caption{BepiColombo ground track during the 2nd Venus flyby. Terminator and illuminated regions are referred to the flyby starting epoch. Epochs are expressed in UTC.}
    \label{fig:VGT}
\end{figure}

\begin{figure}[!ht]
    \centering
    \subfloat[\centering XY VSO frame]{{\includegraphics[width=0.45\textwidth]{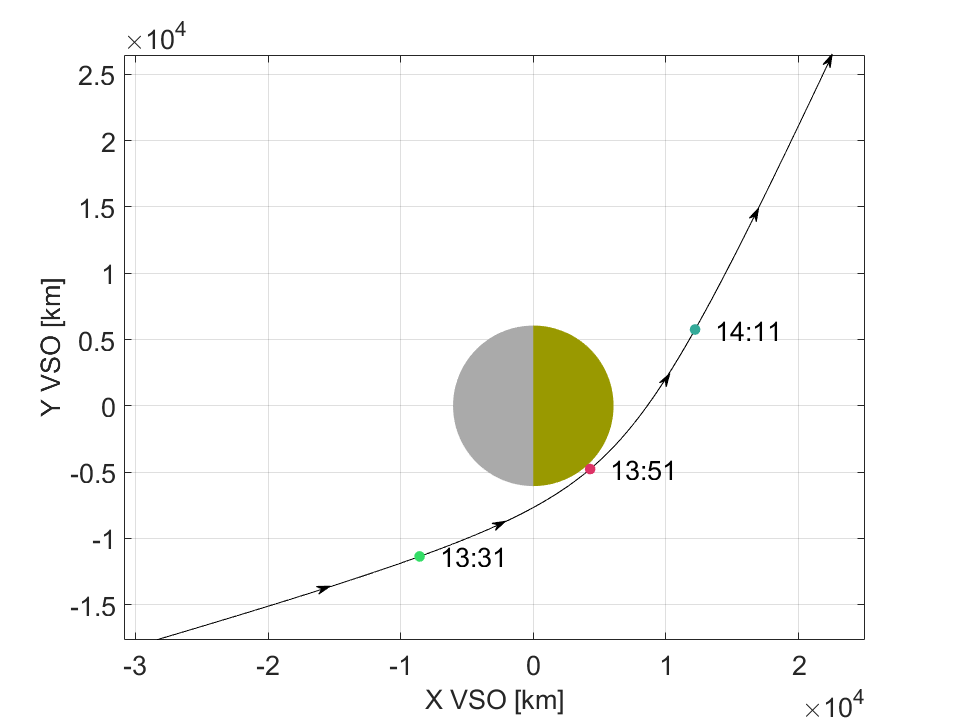} }}%
    \qquad
    \subfloat[\centering XZ VSO frame]{{\includegraphics[width=0.45\textwidth]{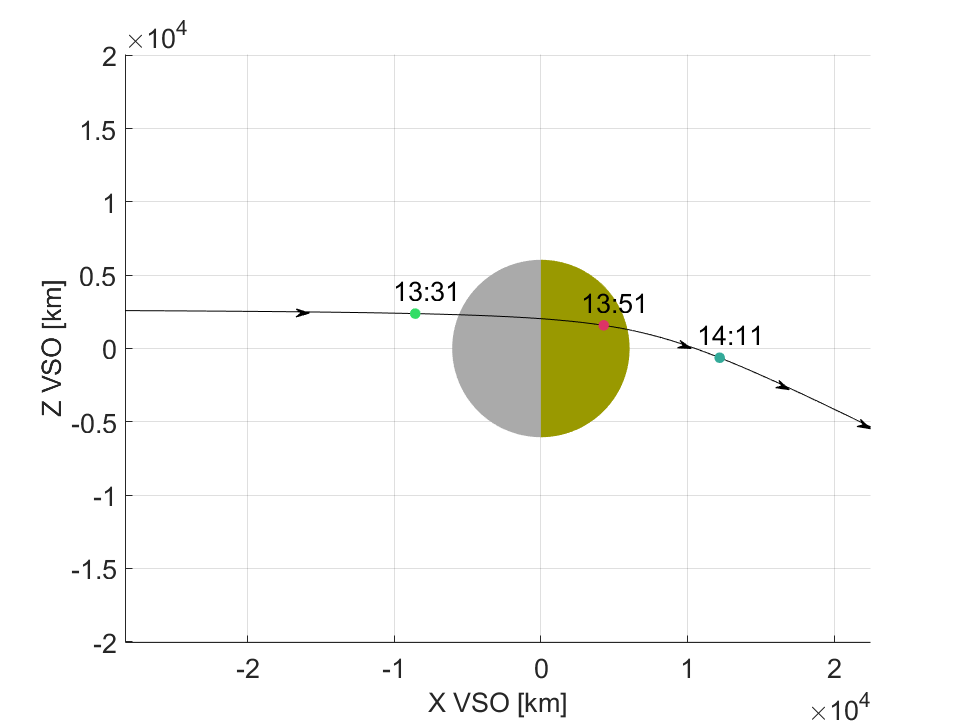} }}%
    \caption{BepiColombo 2nd Venus flyby geometry in VSO reference frame. Red dot references the BepiColombo closest approach. Epochs are expressed in UTC. }%
    \label{fig:MPOVG}%
\end{figure}

For its characteristic, clearly shown in Figure \ref{fig:VGT} and figure \ref{fig:MPOVG}, Venus Swing-by\#2 (VSB2) was highlighted as a relevant event for many instruments to monitoring the Venus environment, the Venusian exosphere ions abundance and the crossing of magnetosphere boundaries.\\


An accurate analysis of spacecraft attitude is the first step to understand the forces acting on board.
The Venus Solar Orbital frame (VSO) is a dynamic frame defined as a reference frame centred on Venus, +X axis pointing at the Sun and +Z directed through the normal to the orbital plane (positive North), so it is very useful to easily identify the position of the Sun. From  figure \ref{fig:MPOVG} a) it is easy to assess that no eclipse events were present during this flyby.

\begin{figure}[!ht]
\centering
    \includegraphics[width=0.5\textwidth]{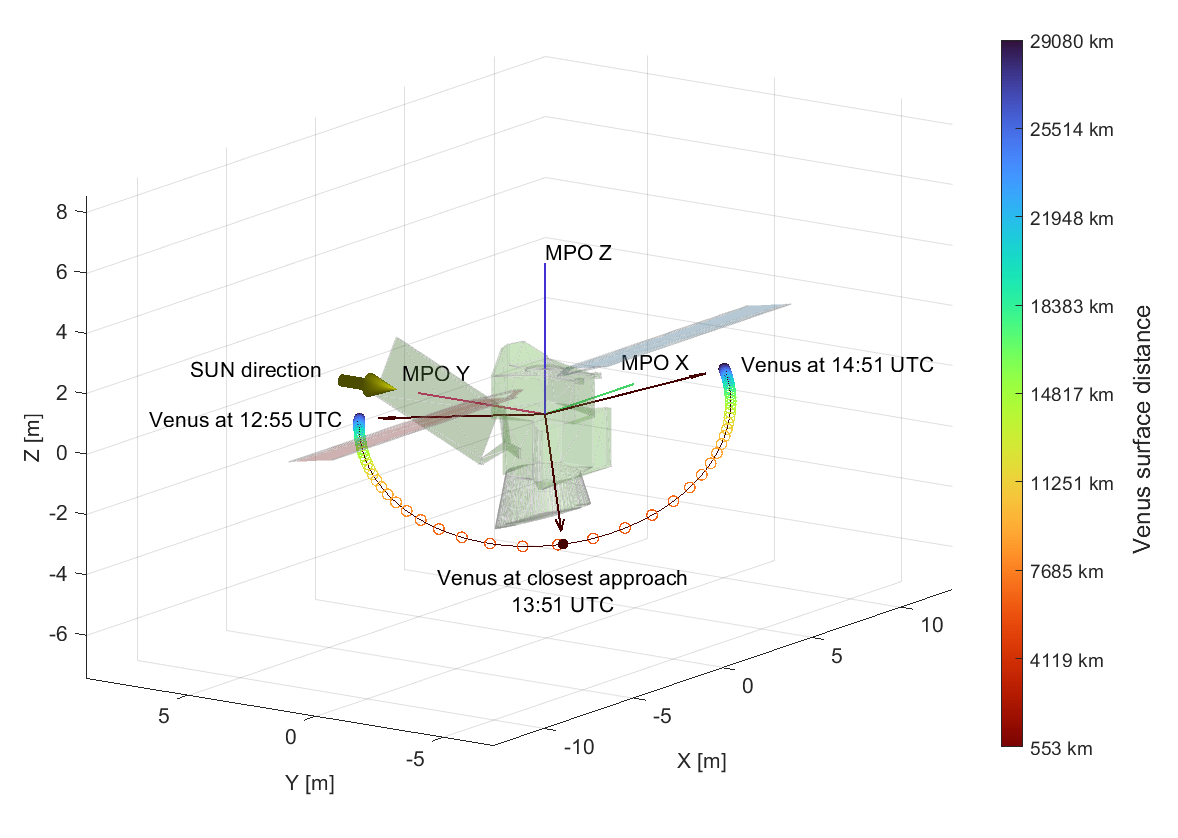}
    \caption{Venus position in MPO body frame during the entire flyby. The color-scale bar refers to the BepiColombo altitude over the Venus surface.}
    \label{fig:MCS_Venus_Attitude}
\end{figure}

The ISA team used the SPICE kernels to reconstruct the attitude of the probe (\citealp{ACTON20189, ACTON199665, SPICE}). To describe the orientation of MCS in space and the forces acting on it, a body fixed reference frame was used. The reference is defined as follows: 
$\hat{\boldsymbol{{Z}}}_{MPO}$ it is along the axis of the MCS, directed as the the MTM four SEP thrusters, the $\hat{\boldsymbol{{Y}}}_{MPO}$ axis is the opposite to the normal to the MPO radiator and, completing the right-handed frame, $\hat{\boldsymbol{{X}}}_{MPO}$ axis points along the MTM Solar Arrays axis. 
Figure \ref{fig:MCS_Venus_Attitude}, produced with a tool developed by ISA team to visualise the MCS attitude and appendices position during the cruise, shows the described body fixed MPO frame (coincident with MCS reference)  and the direction of Venus during the VSB2.  
The spacecraft was commanded to be quasi-inertial during the $t = CA \pm 1hrs $ period, so the Sun Aspect Angle (SAA) did not vary during this phase and the $\hat{\boldsymbol{{Y}}}_{MPO}$ axis was constantly kept pointing the Sun as it happens during almost all the cruise.
On the other hand, Venus can be seen to cover more than 180 degrees in azimuth and forty degrees in elevation in MCS body fixed reference frame. In the figure ( \ref{fig:MCS_Venus_Attitude}) the altitude of the spacecraft over the Venus surface is also reported in the colorscale.

\section{ISA measurement campaign during the second Venus swing-by (VSB2) }

ISA approached the cruise phase with the intent to acquire as much information possible on the behaviour of the instrument, so the team was one of the first to identify the VSB2 as an important occasion to get precious data to be used for internal characterisation scopes by comparing its measurements with the expected MCS NGPs and tidal effects.
Hence, to avoid any spacecraft-generated disturbances, ISA team requested to ESOC operations to maintain the spacecraft in quasi-inertial pointing, to not perform reaction wheels desaturation manoeuvres and to avoid appendages movements, including HGA pointing mechanism to be triggered in hold status for the period $t = CA \pm 1hrs $. 

The unit was operated for the VSB2 to approach Venus in thermalised conditions and in nominal gain (ensuring an electronic noise floor at $10^-9\ m/s^2/\sqrt{Hz}$) and at nominal acquisition rate (Observation mode 10\ Hz). Furthermore, Housekeeping data, shows stable temperatures of ISA sensing elements, despite the closeness of  Venus during the swing-by provoked a long trend term heat of the ISA mounting panel.

\subsection{Expected accelerations: characterisation and order of magnitude}

Here after are reported, divided by type, the expected accelerations order of magnitude during the VSB2.
Table \ref{tab:expectedAcc} gives a quick summary while in the next paragraphs references of how they are calculated is given. 

\begin{table}[h]
    \centering
    \small
        \begin{tabular}{m{6em} m{8em} m{5em}}
            \toprule
            Acceleration Type & Acceleration Source & expected value $[m/s^2]$\\
             \toprule
             Conservative &  Gravity Gradient & $1.12*10^{-6}$\\
            \toprule
            \multirow{4}{2em}{ NGPs } & SRP & \textit{neg.}\\
             &Planetary Albedo & $3.6*10^{-8}$\\
             &Planetary IR & $2.5*10^{-8}$\\
             &Thermal Recoil & \textit{neg.}\\
             \toprule
        \end{tabular}
        \caption{Expected accelerations: characterisation and order of magnitude.}
        \label{tab:expectedAcc}

\end{table}

\subsubsection{Expected Gravity Gradients}

Main scope of the ISA planning for the VSB2 was to measure its first internal physical signal, the Gravity Gradient. Indeed, even before to know the spacecraft commanded attitude, a rough calculation of the Gravity Gradient expected effect ($a_{GG}$) between the MCS CoM and ISA mounting point, can be done starting from the Venus Standard Gravitational Constant $\mu_{\mathVenus} = 3.2486*10^5 [km^3/s^2]$ and assuming a $R_{alt} = 550 km$ altitude passage ($R =  6,051 Km + R_{alt} $) with a 1 m base displacement from the CoM: \[a_{GG}=\frac{\mu_{\mathVenus}}{R^3}*1=1.12*10^{-6} [m/s^2],\]so, well over the ISA noise floor. 

\subsubsection{Expected non-gravitational perturbations}
\paragraph{Solar radiation pressure variation}
As previous reported, the SAA of the spacecraft do not change during the flyby, neither Sun occultations are encountered in the trajectory close to Venus. As a consequence, the direct Sun irradiance do not change significantly in the period under analysis. Furthermore, the MCS appendices, including the MTM Solar Arrays, are commanded to be fixed for the $CA \pm 1hrs $. The conclusion is that the  Solar Radiation Pressure variation (SRP) could be considered negligible during the whole flyby. 

\paragraph{Planetary albedo} \label{PVR}
From figure (\ref{fig:MPOVG}) can be observed that the spacecraft approached Venus from its night side. Indeed, from some minutes before the closest approach, the spacecraft crossed the Venus terminator and, gradually, the entire illuminated disk of the planet was exposed to the spacecraft.
An upper order of magnitude of such an effect can be calculated considering the fact that, at closest approach, almost all the full disk of Venus is illuminated. 
A raw formula can be used here as the full Venus disk is 550 km far from the spacecraft:
\[ NGP_{albedo}=\frac{\overline{W}_{alb}}{c}*\frac{S_{SC}}{m_{SC}},\]
where, $c$ is the speed of light, $S_{SC}$ is the spacecraft active surface, $m_{SC}$ is the spacecraft mass and ${\overline{W}_{alb}}$ is the Venus albedo irradiance, calculated taking into account the $\theta$ angle, i.e. the half angle under which Venus is observed by the spacecraft at CA altitude 550 km. Assuming that, from the BepiColombo ESA internal documentation, the MCS assorbance  $\alpha_{SC} = 0.78$ and the albedo optical coefficient of the planet Venus $alb = 0.75$ from \cite{1983vens.book...45M}, the $\overline{W}_{alb}$ can be arranged as:  
\begin{equation}
\label{eq:albedo_irradiance}
{\overline{W}_{alb}}=\frac{2\pi\left( 1-\cos\left(\theta\right)\right)*\left(1+\alpha_{SC}\right)*alb}{4\pi}*\overline{W}_{\mathVenus}
\end{equation}

Since the Solar constant at Venus during the BepiColombo flyby ($0.72 AU$) can be approximated as: \[\overline{W}_{\mathVenus}=\overline{W}_{\mathEarth}\left(\frac{R_{\mathVenus}}{R_{\mathEarth}}\right)^2 = 2670  [W/m^2]   \] 

and considering the spacecraft maximum possible surface exposed to Sun about $S_{SC} = 37 m^2 $, the spacecraft mass $m_{SC} = 3991 kg $, hence the NGP due to albedo can be evaluated as:
\[ NGP_{albedo} =    3.6*10^{-8}  [m/s^2]
\]

Hence, the upper approximation of the albedo radiation acceleration expected on the spacecraft is about the order of magnitude of the nominal ISA intrinsic noise ($1*10^{-8} [m/s^2/sqrt(Hz)]$). 

\paragraph{Planetary Infrared radiation} \label{PIR}
Venus atmosphere can be considered as a black body emitting at about $240\ K$. The corresponding average body irradiance can be estimated in $154\ W/m^2$. 

Applying similar considerations used for the Planetary Albedo in equation \ref{eq:albedo_irradiance}, we can estimate the maximum infrared acceleration due to the radiation emitted by the planet in wave number between $230$ $cm^{-1}$ and $2300$ $cm^{-1}$, \cite{TAGUCHI2012502}, as:

 \[ NGP_{IR} = 2.5*10^{-8}  [m/s^2] \]

\paragraph{Thermal recoil acceleration} \label{TRP}

A Thermal Recoil Acceleration (TRA) is due to  unbalanced thermal emissions of spacecraft surfaces, generating a net force pushing the spacecraft body in the opposite direction. 
A good estimate of that force can be carried out by collecting the temperature of the body surfaces and calculate the emission for each surface. In our study it is important to calculate an order of magnitude of the effect, to be sure that the acceleration source shall be neglected when cleaning the data. 
As already mentioned, the spacecraft points its $\hat{\boldsymbol{{Y}}}_{MPO}$ towards the Sun, so, MCS surfaces oriented along the +Y axis can be considered at a temperature much greater that the surfaces pointing -Y. 
During the cruise, at a Sun-spacecraft distance comparable with the Venus orbit, the delta average temperature between the two spacecraft sides can be kept around $120\ K$. Looking at thermistors onboard the spacecraft, we knew that the Venus albedo heated the -Y side of the MCS, increasing the temperature of of about $30\ K$.
In terms of acceleration, considering that ISA is a relative accelerometer, the changes of acceleration in its frequency band can be only addressed to the $30\ K$ variation, since the +Y side of the MCS did not change significantly its temperature. We can conclude that this effect is well below the instrument sensitivity.

\subsection{Gravity Gradients measurement}

Figure \ref{fig:VFBISA} shows the calibrated acceleration detected by ISA (solid lines) over-imposed to the expected calculated accelerations (dotted lines) that, as expressed in the previous paragraph, can be considered mainly the gravity gradients signals expected at each ISA sensing elements. 
ISA team, by got the executed attitude and the most accurate final trajectory using the BepiColombo operational SPICE kernels,precisely simulated the expected Gravity Gradient signal experienced by each accelerometer sensing elements. The equation for the Venus field monopole effect is used:

\begin{equation}
\textbf{a}^j_{GG}=\frac{\mu_{\mathVenus}}{|\textbf{R}_0|^3}\bigg(3\hat{\textbf{R}}_0\hat{\textbf{R}}_0^T-\textbf{I}_{3\times3}\bigg)\textbf{r}^j_{ISA},
\label{aGG}
\end{equation}

where $\mu_{\mathVenus} = GM_{\mathVenus}$ is the Venus standard gravitational coefficient,  $\textbf{R}_0$ is the vector linking the CoM of MCS with the Venus center, $\hat{\textbf{R}}_0$ is its unit vector, the $\textbf{I}_{3\times3}$ is a 3x3 identity matrix and $\textbf{r}^j_{ISA}$ is the vector of the j-nth accelerometer (Acc0 = +X ISA directed, Acc1 = Y ISA directed, Acc2 = Z ISA directed) from the MCS CoM. The CoM position used in our simulations has been taken from the ESA/ESOC manoeuvre file.\par
The figure \ref{fig:VFBISA} undoubtedly unveils that the simulated Gravity Gradient signal agrees with the ISA measured data.
Hence, for the first time,  a direct measurement of the Gravity gradient was carried out for an extraterrestrial body. 

\subsection{Unexpected Non Gravitational Perturbation}
\begin{figure*}
\centering
    \includegraphics[width=1\textwidth]{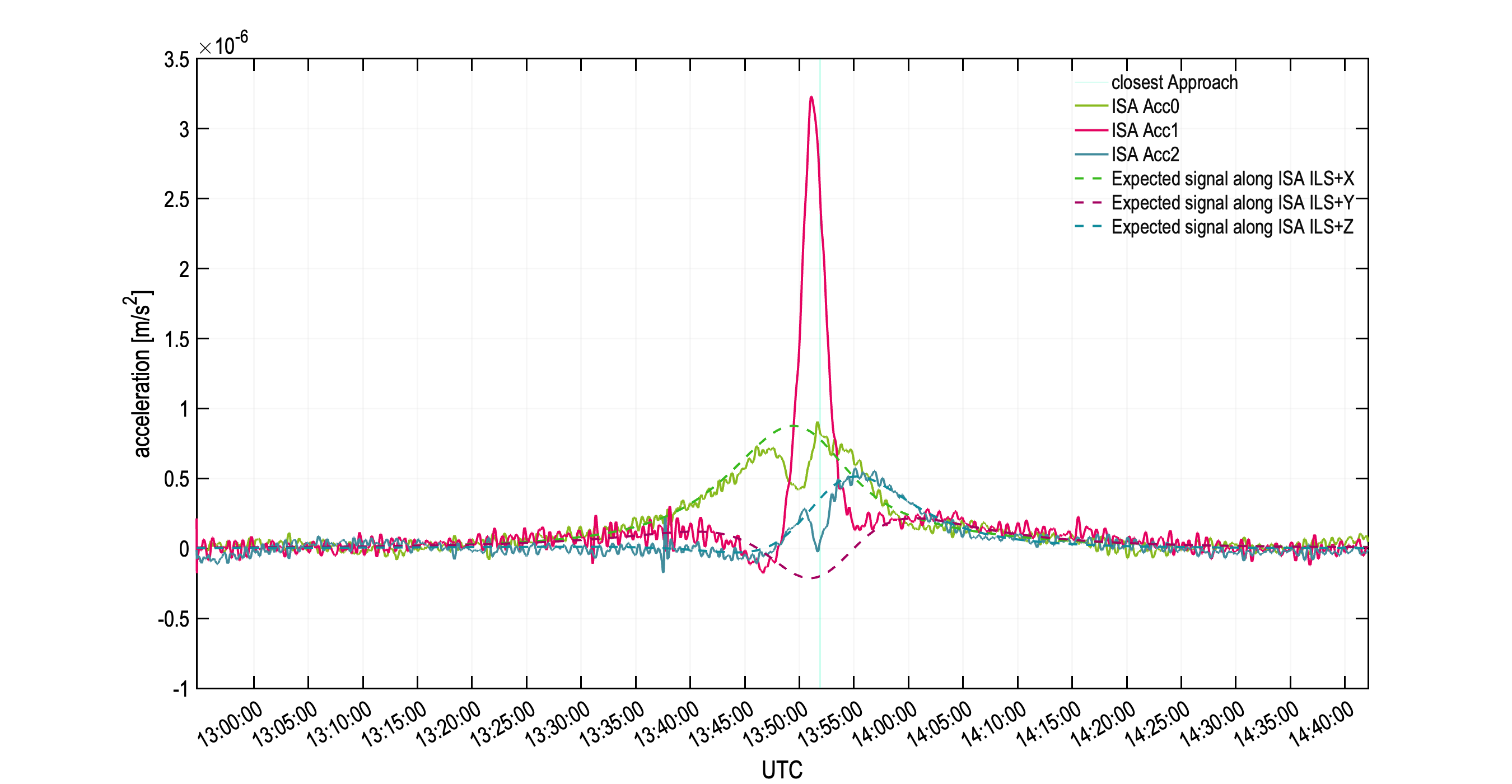}
    \caption{ISA measured calibrated accelerations during the second Venus flyby.}
    \label{fig:VFBISA}
\end{figure*}

The figure (\ref{fig:VFBISA}) also shows that in the period from $CA-4'$ to $CA+3'$ a spike acceleration, magnified along Acc1 (Y ISA directed and hence very close to Y MCS), but visible in all the channels, caused the measurement to deviate, by order of magnitude, from the expected signals. 
Occurring at the CA, the spike appeared in coincidence with the direct albedo acceleration and the Planetary Infrared radiation pressure expected maximum,  making impossible to distinguish them among data. 

This spike soon was considered completely unexpected because it did not match in amplitude and behaviour to any of the predicted accelerations listed in the previous section. At this point, ISA team wondered if this signal was a real physical acceleration acting on the spacecraft or an artifact introduced by the instrument.\par

\subsubsection{On board net torque analysis}
To understand if ISA measured an artifact or the whole spacecraft experienced a real external acceleration, the team looked for in the Attitude-Orbit-Control-System (AOCS) data for signatures indicating a torque caused by an external force acting on the MCS. Since, as described in \ref{BC_VSB_2}, the commanded attitude was quasi-inertial and the +Y Axis always pointed towards the Sun, the main torque compensation of the AOCS should have been the gravity gradient torque. So, linking the spacecraft moment of inertia, trajectory and attitude, the expected gravity gradient torque was calculated using the following equation:

\begin{equation}
\boldsymbol{T}_{GG}=3n^2\hat{\boldsymbol{R}}_0^B\times\bigg(J\hat{\boldsymbol{R}}_0^B\bigg)
\label{torqueGG}
\end{equation}

where $n$ is the spacecraft mean motion, $J$ is the spacecraft moment of inertia and $\hat{\boldsymbol{R}}_0^B$ is the Venus-spacecraft direction.\par
\begin{figure}[!ht]
\centering
    \includegraphics[width=0.5\textwidth]{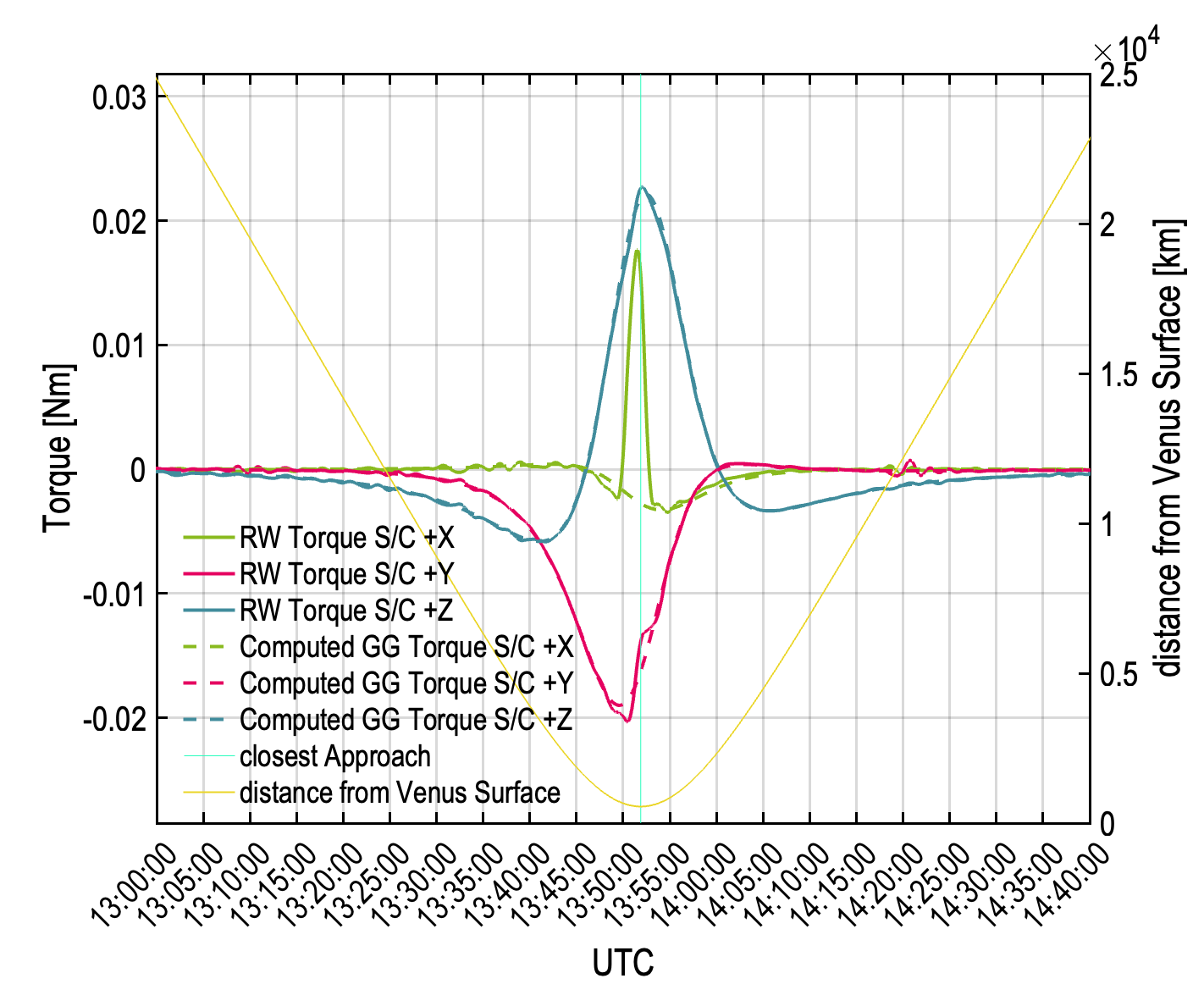}
    \caption{Comparison between reaction wheel executed torque and computed signal due to gravity gradient disturbance torque.}
    \label{fig:VFBtorque}
\end{figure}

Figure \ref{fig:VFBtorque} shows the gravity gradient estimated torques (dashed lines) around the MCS body axis, computed using the equation above, over-imposed to the torque effectively commanded by the AOCS (solid lines). It is clear that the curves are compatible everywhere except for the time span [13:49:30 UTC, 13:53:20 UTC] = $[CA-2', CA+1']$ when the unexpected acceleration spike, described in the previous paragraph, was detected. The misalignment is visible on the three axes, but is clearly highlighted on the $MCS +X$ axis.\par
This circumstance let the ISA team argue that exactly at the time of the spike in acceleration detected by ISA, BepiColombo AOCS compensated a torque to keep the spacecraft at the commanded attitude, mainly around the S/C+$\hat{\boldsymbol{{x}}}_{sc}$ axis: direction fully compatible with a force that pushed the spacecraft mainly along +$\hat{\boldsymbol{{y}}}_{sc}$ direction.

If the disturbance torque detected by the AOCS is caused by the acceleration detected by ISA, the angle between the force direction and the torque direction must be as close as possible to 90 degrees. Indeed, the relationship between these quantities can be written as:

\begin{equation}
\boldsymbol{T}_{dis}=m_{S/C}\boldsymbol{r}_A\times\boldsymbol{a}_{ISA}^{s}
\label{linconstr}
\end{equation}

where $\boldsymbol{T}_{dis} = [{T}_{x,dis},{T}_{y,dis}, {T}_{z, dis}]$ is the difference between the total torque exerted on the spacecraft by the reaction wheel $\boldsymbol{T_{RW}}$ and the expected torque due to the gravity gradients $\boldsymbol{T_{GG}}$ expressed in the body fixed reference frame. The quantity $\boldsymbol{r}_A = [x_A,y_A,z_A]$ is the vector from the S/C COM to the force application point expressed in the body fixed reference frame and $\boldsymbol{a}_{ISA}^{s} = [a_{x,ISA}^s,a_{y,ISA}^s,a_{z,ISA}^s]$ is the spike acceleration measured by ISA computed as the difference between the solid and the dotted lines of figure \ref{fig:VFBISA}.\par
The misplacement from the orthogonality between $\boldsymbol{T}_{dis}$ and $\boldsymbol{a}_{ISA}^s$ can be computed as the quantity $\beta$:

\[\beta = \arcsin(\frac{\boldsymbol{T}_{dis}\cdot\boldsymbol{a}_{ISA}^s}{\abs{\boldsymbol{T}_{dis}}\abs{\boldsymbol{a}_{ISA}^s}})\]. 

\begin{figure}[!ht]
\centering
    \includegraphics[width=0.5\textwidth]{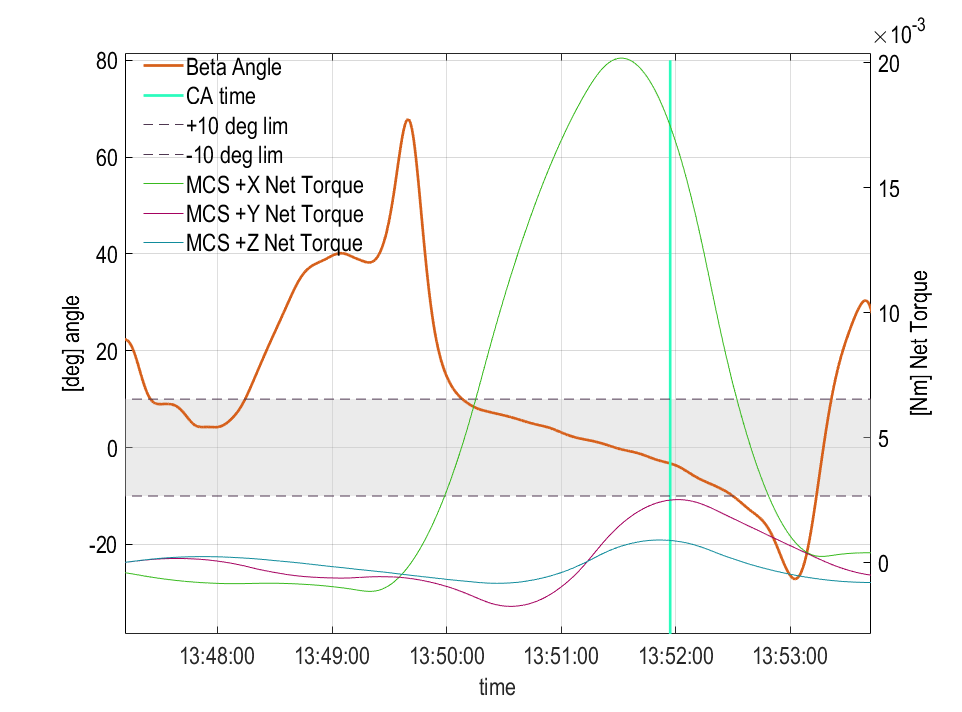}
    \caption{Angle between the measured disturbance torque and the measured ISA acceleration.}
    \label{fig:beta_angle}
\end{figure}

From the figure  \ref{fig:beta_angle} it can be observed that when the disturbance torque starts to act on the spacecraft, the $\beta$ angle reaches the expected value of 0 deg, proving that the AOCS and ISA observations are related to the same physical phenomenon. 

The fact that two independent sensors measured the same event gave ISA team the opportunity to use the equation \ref{linconstr}, linking the two quantities, to resolve for the unknown force application point vector $\boldsymbol{r}_A$ in the period in which $\beta$ angle is close to $0$ degrees.
ISA team used a non-linear least-square estimation method to minimise the function:

\begin{equation}    
{\varepsilon } =  \sum_{i=1}^{N}{e_i^2} = \sum_{i=1}^{N}\left({\textbf{ {T}}_{dis}}|_{i}^{C} - {\textbf{ {T}}_{dis}}|_{i}\right)^2,
\label{nleastsquare}
\end{equation}

where the $i-th$ residual $e_{i}$ at at time $t_{i}$ is calculated as the difference between ${\textbf{ {T}}_{dis}}|_{i}^{C}$, computed following the equation \ref{linconstr}, and ${\textbf{ {T}}_{dis}}|_{i}$ effectively measured from AOCS at time $t_{i}$ but cleaned from the Gravity Gradient effects.
The estimation is performed every $N$ acceleration samples to follow the evolution of the application point $\boldsymbol{r}_A$ during the time.\par
\begin{figure}[!ht]\centering
    \subfloat[]{
        \label{fig:Force_application_point_in_MCS_frame_a}
        \includegraphics[width=.45\linewidth]{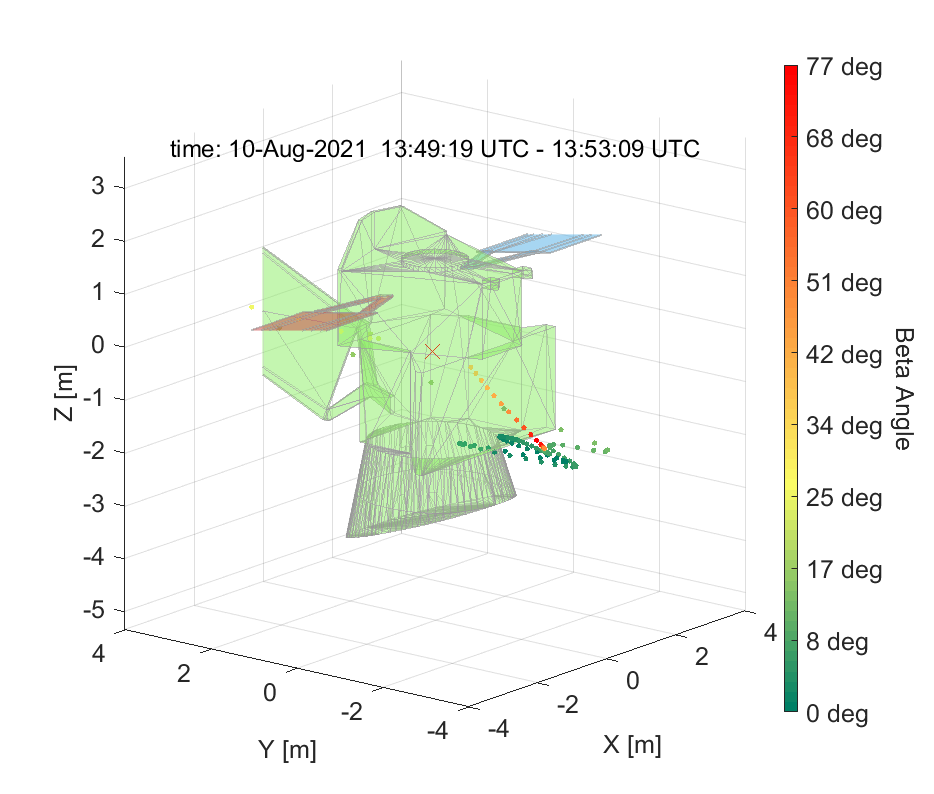}
    }\hfill
    \subfloat[]{
        \label{fig:Force_application_point_in_MCS_frame_b}
        \includegraphics[width=.45\linewidth]{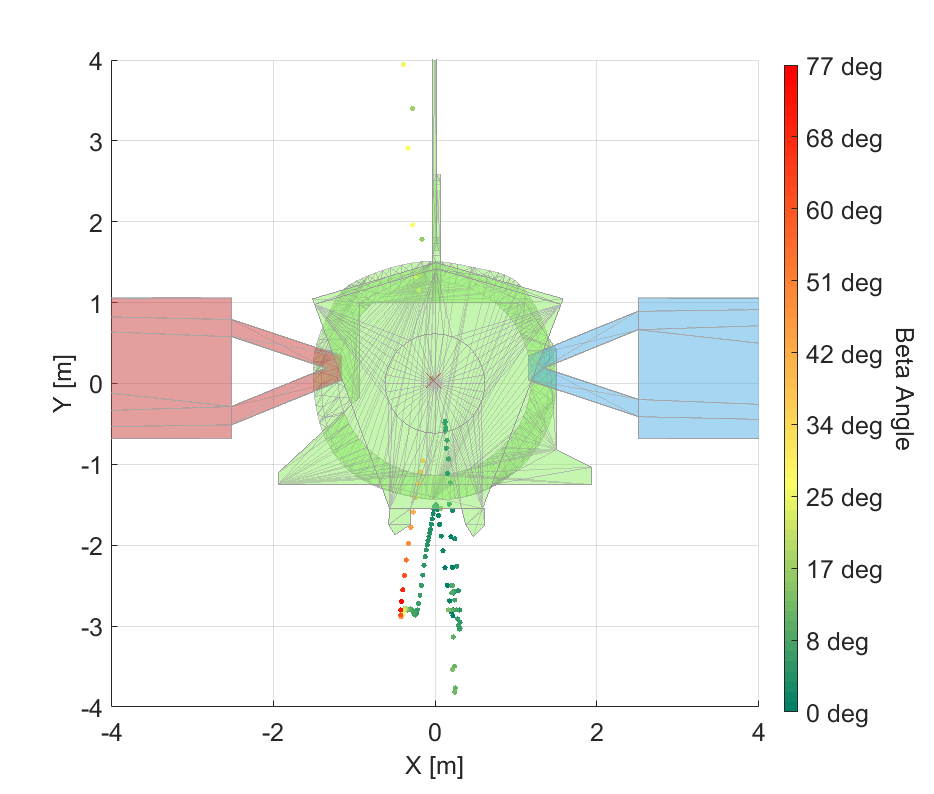}
    }\par
    \subfloat[]{
        \label{fig:Force_application_point_in_MCS_frame_c}
        \includegraphics[width=.45\linewidth]{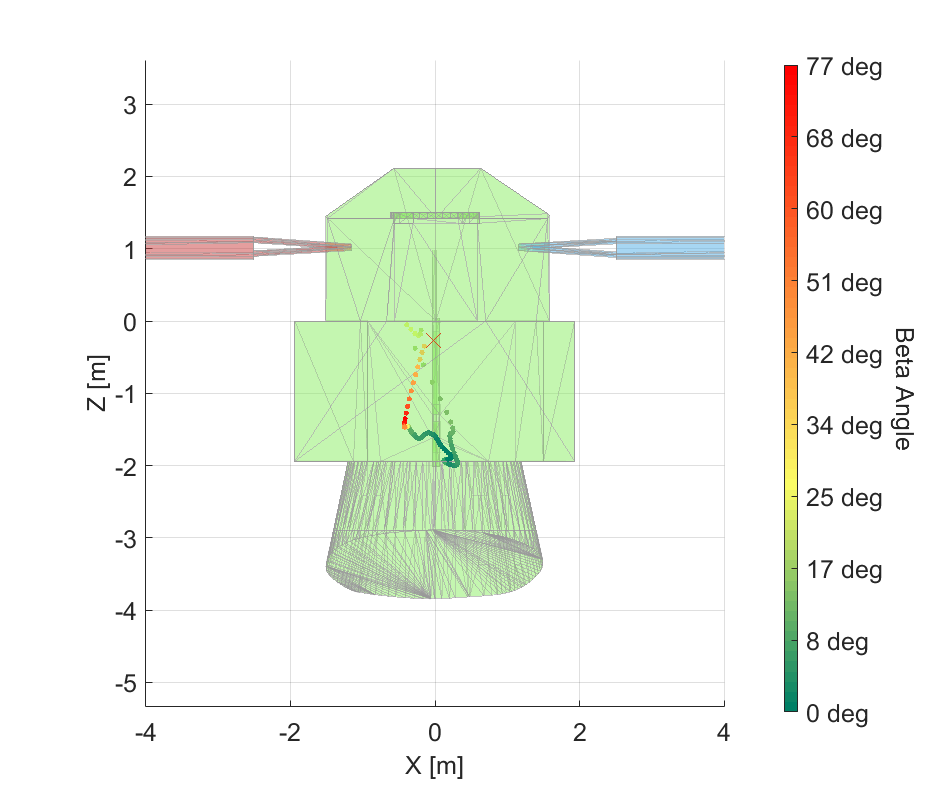}
    }\hfill
    \subfloat[]{
        \label{fig:Force_application_point_in_MCS_frame_d}
        \includegraphics[width=.45\linewidth]{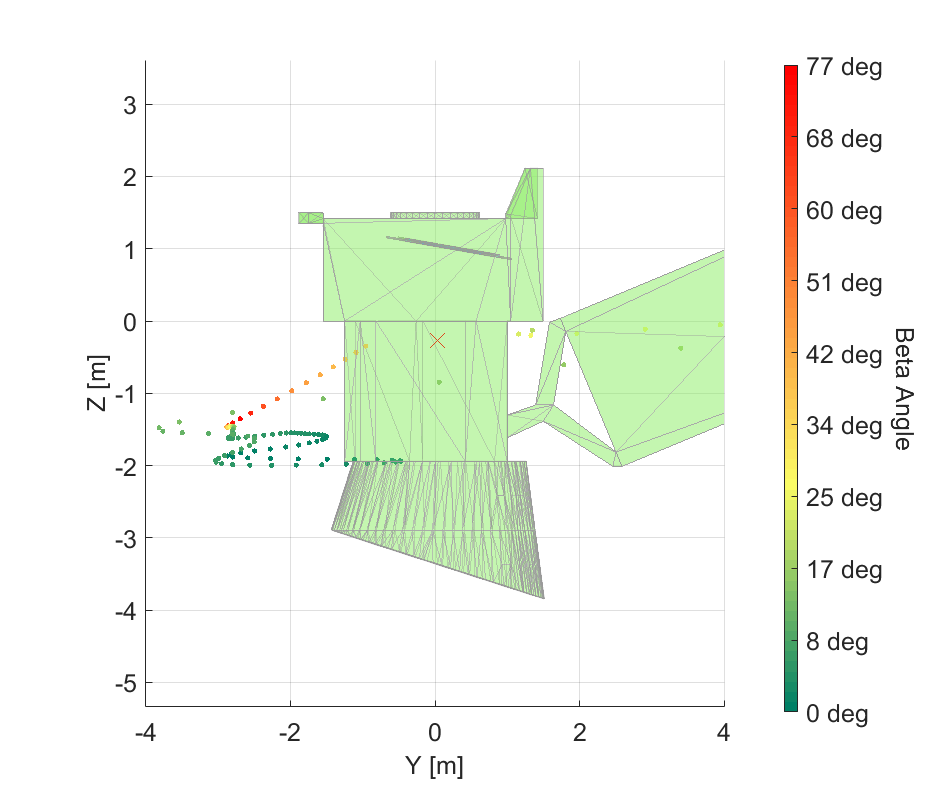}
    }
    \caption{Estimated force actuation points in the MPO reference frame. Dots colors are related to the colormap showing the $|\beta|$ angle.}
    \label{fig:Force_application_point_in_MCS_frame}
\end{figure}

In figures \ref{fig:Force_application_point_in_MCS_frame} are depicted the reconstructed application point as evolving in time, compared with the MCS body model in which markers color are based on the $\beta$ value, the greener are the dots, the closer is the angle to zero.
Despite the estimation solution is not constrained in any direction, it can be noted that green points are all placed in nearby of MPO radiator: the average of the estimated application point, in the interval at low beta angle, is:

\[  [0.0235   -2.1841   -1.7233] [m]\] 

in the MPO reference frame and the $X$ and $Z$ coordinates variation is about few centimeters, when the modulus of $\beta$ angle is under the 10 degrees. However,  $Y$ coordinates shows bigger oscillation in time, moving the application point also outside the MPO body. 

The reason of that is clearly related to the geo\-metry of the reconstructed force direction: when the absolute value of $\beta$ angle is under 10 degrees, the force is almost directed towards the MPO $+Y$, keeping the equation \ref{nleastsquare} bad conditioned for that coordinate.\par
Using the Jacobian matrix, we evaluated from the numerical solution of the non linear least square fit for the equation \ref{nleastsquare}, the standard deviation respect to the converged solution for the three component of the vector. 

\begin{figure}[!ht]
\centering
    \includegraphics[width=0.5\textwidth]{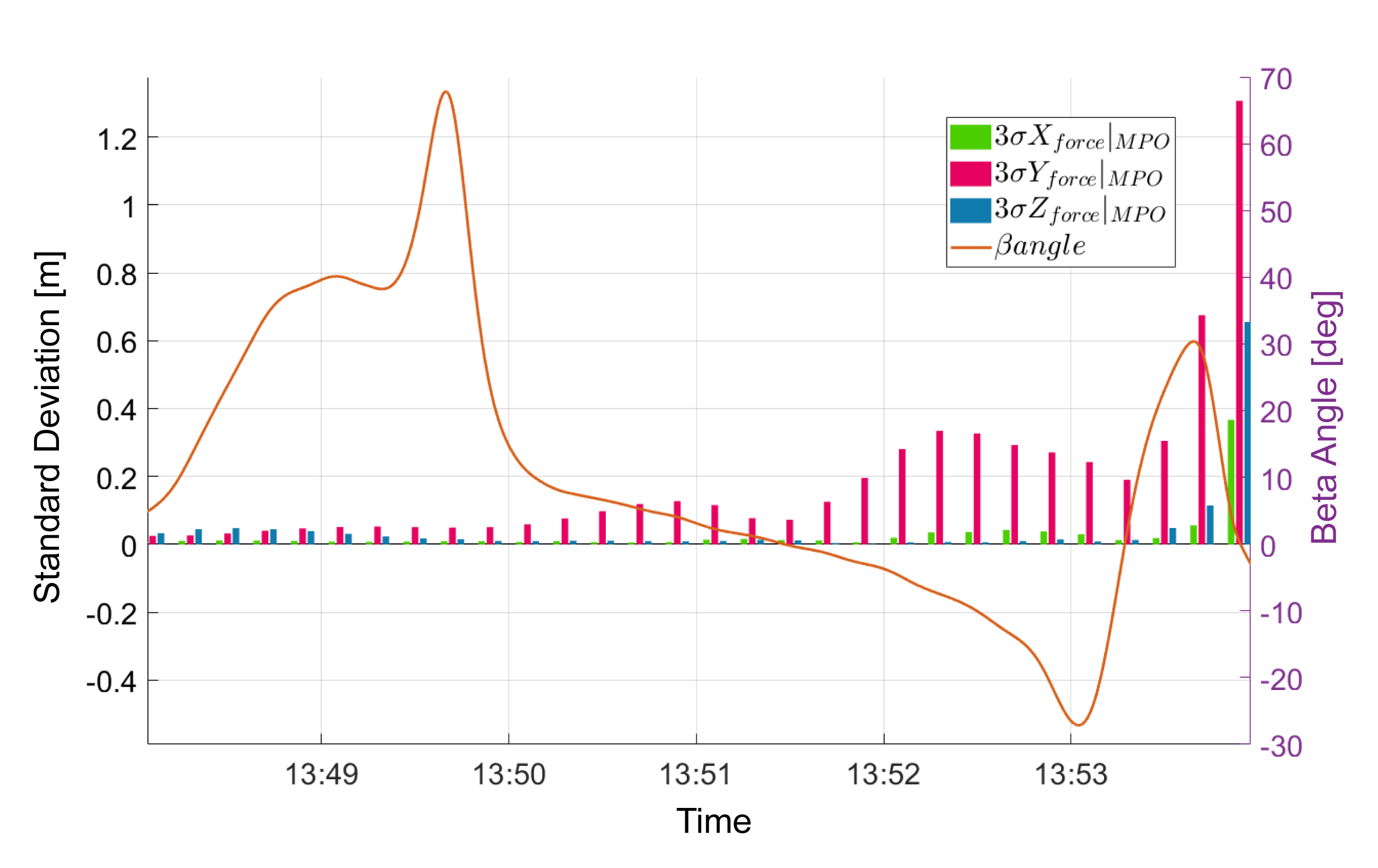}
    \caption{3 sigma formal standard deviation for the three components of the force application point compared with $\beta$ angle}
    \label{fig:sigma}
\end{figure}

Figure \ref{fig:sigma} indicates the sensitivity of the converged solution respect to the variation of the observed data.  As expected, Y coordinate is much more sensitive  with respect to the X and Z, reflecting a bigger uncertainty in the solution. 
The same conclusion can be addressed empirically: if the force would be perfectly along the $Y$ axis, wherever the force is applied on a straight line parallel to this axis it would produce the same torque, letting completely loosing sensitivity on the Y axis.

\subsubsection{Measured Delta V and orbit estimation comparison}
The numerical integration of the ISA measured signal and the expected one, provides the magnitude of the $\Delta V$  on the spacecraft centre of mass due to the unintended acceleration which is $\Delta V = 5.8 \pm 0.4 \times 10^{-4}$ m/s, where the uncertainty has been computed by integrating a sample of data where no signal was expected. 
This value is fully compatible with the analysis based on radio-data, which estimated an impulsive $\Delta V = 5.95 \pm 2.71 \times 10^{-4}$ m/s along +$\hat{\boldsymbol{{y}}}_{sc}$ centred at the peak of the ISA measured spike \cite{DelVecchio10480228}. Previous analysis, made just after the Venus swing-by\#2 by the ESA/ESOC flight dynamics department, proved that the orbit determination software during the swing-by added a stochastic $\Delta V$ close to periapsis with a-priori spherical $\Delta V$ constrained to 0.5  mm/s. The filter estimates a $\Delta V$ along S/C +Y of 0.8 mm/s with a-posteriori sigma of 0.3 mm/s, so moderately observable, but consistent with what ISA measured (ESA private communication, thanks to Frank Budnik, Alkan Altay, Emanuela Bordoni and Daniele Stramaccioni). 

\section{Conclusions}

The presented work analysed the ISA measurements campaign during the second Venus swing-by using a comparison of the acquired data with the expected tidal effects between S/C CoM and ISA sensors, evidencing ISA as the first accelerometer ever that took direct measurements of Gravity Gradients  generated on a spacecraft from an extraterrestrial body.\par
Furthermore, data analysis evidenced an unexpected and almost un-modellable NGP acting on the spacecraft resulting in an acceleration spike of about $3.5\times10^{-6} [m/s^2]$ in a time interval of about 6 minutes around the Closest Approach. The dynamical analysis of ISA data and AOCS data (mainly the RWs torques) allowed to identify the axis and point of application of the force causing such an acceleration. The force was estimated to be applied close to the MPO radiator and the force direction to be almost opposite to the radiator normal, the authors than argue that it was a surface force to push the spacecraft along its +Y axis.
The presence of that unexpected NGP is also supported by the communications that authors had with the MORE team and the ESOC flight dynamic department, from which can be drawn that the orbital trajectory estimation matches with very good accuracy the ISA accelerations measurements, in terms of $ \Delta V $ transferred to the BepiColombo probe exit trajectory (ESOC private communication).
The nature of such a force, and in particular the hypothesis of an out-gassing phenomenon, is deeply investigated in the companion paper of this work (U. De Filippis, "Outgassing event detected by the Italian Spring Accelerometer during the second Venus flyby of BepiColombo").\par
It is worthwhile to remark that the analysis of VSB2 data confirms that ISA like high-sensitivity accelerometers are key instruments for the spacecraft fine-dynamic and Precise Orbit reconstruction in the regime of small and very small accelerations, overcoming the need of modelling them that would be an extremely challenging and sometimes not feasible activity in such complex scenarios.

\section*{Acknowledgements}
The present work has been carried out in cooperation and with the support of the Italian Space Agency – ASI (under cooperation agreement n. 2017-47-H.0) and European Space Agency – ESA (under contract No. 4000119356/16/ES/JD).

\bibliographystyle{plainnat} 
\bibliography{references, references_b}

\end{document}